\documentclass[a4paper,11pt]{article}
\usepackage{pos}
\usepackage{amsmath}

\title{Hunting for bumps in the diffuse high-energy neutrino flux}
 \ShortTitle{Bump-hunting in diffuse neutrinos}

\author*[a]{Damiano F. G. Fiorillo}
\author[a]{Mauricio Bustamante}

\affiliation[a]{Niels Bohr International Academy, Niels Bohr Institute,\\University of Copenhagen, 2100 Copenhagen, Denmark}

\emailAdd{damiano.fiorillo@nbi.ku.dk}
\emailAdd{mbustamante@nbi.ku.dk}

\abstract{The origin of the TeV--PeV astrophysical neutrinos seen by the IceCube telescope is unknown. If they are made in proton-photon interactions in astrophysical sources, their spectrum may show bump-like features. We search for such features in the 7.5-years High-Energy Starting Events (HESE), and forecast the power of such searches using larger data samples expected from upcoming telescopes. Present-day data reveals no evidence of bump-like features, which allows us to constrain candidate populations of photohadronic neutrino sources. Near-future forecasts show promising potential for stringent constraints or decisive discovery of bump-like features. Our results provide new insight into the origins of high-energy astrophysical neutrinos, complementing those from point-source searches.}

\ConferenceLogo{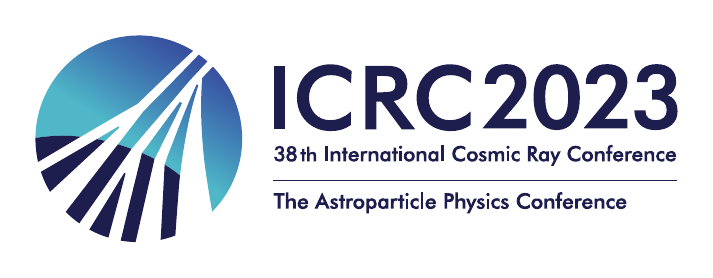}

\FullConference{%
38th International Cosmic Ray Conference (ICRC2023)\\
  26 July - 3 August, 2023\\
  Nagoya, Japan}

\begin{document}
\maketitle

\section{Introduction}\label{sec:intro}

The IceCube neutrino telescope has detected a diffuse flux of high-energy neutrinos in the range between $10$~TeV and $10$~PeV. The origin of these neutrinos is so far unknown. The bulk of them is likely produced in extragalactic sources capable of accelerating cosmic rays (CRs) to EeV-scale energies. However, whether a single or multiple populations of sources contribute sizably to this flux is not clear. One strategy to make progress on this question is the identification of individual point sources belonging to specific source classes. A few possible transient sources have been identified, including the blazar TXS 0506+056 and three tidal disruption events, and has recently claimed the first detection of a steady neutrino source, the Seyfert galaxy NGC 1068. However, even these discoveries have not provided a definite answer to the question of the origin of the diffuse flux.

A complementary strategy, which we adopt here, is to examine the spectrum of the diffuse neutrino flux. Our aim in Ref.~\cite{Fiorillo:2022rft}, where we give full details, is to test whether spectral features coming from the superposition of more than one source population could be visible, given a sufficiently accurate measurement of the neutrino spectrum. We consider two broad classes of candidate high-energy neutrino sources: those where neutrinos are made primarily in cosmic-ray interactions with ambient matter---i.e., proton-proton ($pp$) sources---and those where neutrinos are made primarily in cosmic-ray interactions with ambient radiation---i.e., photohadronic ($p\gamma$) sources.  In both, neutrinos come from the decay of the short-lived particles---pions and muons, mostly---born from these interactions.  However,  they emit neutrinos with different energy spectra.

Neutrinos from $pp$ sources have a power-law spectrum, inherited from their parent cosmic rays. Candidate $pp$ sources include starburst galaxies, galaxy clusters, and low-luminosity active galactic nuclei (AGN).  Neutrinos from $p\gamma$ sources have instead a ``bump-like'' spectrum, centered around an energy determined by the properties of the interacting photons and cosmic rays.  Candidate $p\gamma$ sources include gamma-ray bursts, low-luminosity AGN, radio-quiet AGN, radio-loud AGN, BL Lacertae AGN, flat-spectrum radio quasars, and tidal disruption events. In Ref.~\cite{Fiorillo:2022rft} we include a list of references detailing the production mechanisms in these sources.

The classification into $pp$ and $p\gamma$ is admittedly only schematic. Most candidate source classes can in reality produce neutrinos via both mechanisms. Furthermore, $pp$ sources with a particularly hard parent cosmic-ray spectrum may give rise to bump-like spectra, so there is not a one-to-one correspondence between the neutrino spectral features and the neutrino production mechanism. Also, we do not test predictions of specific source models, but the presence of generic spectral features due to $pp$ and $p\gamma$ production.

The bulk of the present-day IceCube data are well described by a pure-power-law spectrum, i.e., $\Phi_\nu \propto E^{-\gamma}$, with $\gamma \approx 2.37$--2.87~\cite{IceCube:2020wum}, depending on the data used.  However, the large present-day uncertainty in the measured energy spectrum might be hiding deviations from a pure power law.  To wit, while there is no marked preference for alternatives to a pure power law, they are not strongly disfavored~\cite{IceCube:2020wum}.  More complex possibilities are allowed too, e.g., a two-component model with $pp$ sources dominating up to PeV energies and $p\gamma$ sources dominating above~\cite{Palladino:2018evm, Ambrosone:2020evo}, or $p\gamma$ sources opaque to gamma rays~\cite{Capanema:2020rjj, Capanema:2020oet} dominating below 60~TeV~\cite{Murase:2015xka}.

Motivated by these works, we search for the presence of power-law and bump-like diffuse flux components in present-day IceCube data, and make near-future forecasts using the combined exposure of upcoming neutrino telescopes. For our present-day search, we use the most recent data release of the 7.5-year High Energy Starting Events (HESE)~\cite{IceCube:2020wum}. For our forecasts, we assume that upcoming telescopes will have comparable efficiency to IceCube HESE detection. Rather than using specific source models, we use flexible parameterizations for the power-law and the bump-like features we aim to test.

We address two questions. First, we show that with the exposure of upcoming telescopes we may soon distinguish decisively between a single-component ($pp$ only {\it or} $p\gamma$ only) and a multi-component ($pp$ {\it and} $p\gamma$) description of the diffuse neutrino flux. Second, we show that, if no bump-like feature will be observed, we can still constrain the size of the neutrino flux of $p\gamma$ sources, with arguments that are independent from point source searches.

\section{Power-law and bump flux components}\label{sec:pl_bump_components}

The crux of this work is that different sources produce different neutrino spectra. In $pp$ sources, since neutrinos carry a fixed fraction of the parent proton energy, the neutrino spectrum simply mimics the CR spectrum. Since in most acceleration mechanisms one expects CR power-law spectra $dN/dE\propto E^{-\gamma}$ with $\gamma\gtrsim 2$, e.g., in diffusive shock acceleration, $pp$ neutrinos typically have a soft power-law spectrum, with a spectral index larger than 2. However, this is not a general conclusion, and if harder spectra are attained, either due to acceleration mechanism or energy-dependent escape, the conclusions may differ. 

In $p\gamma$ sources, the spectrum does not directly mimic the CR spectrum, but is instead hardened because of the larger efficiency of resonant proton-photon scattering at high energies (see, e.g., Ref.~\cite{Fiorillo:2021hty} and references therein). Therefore, the neutrino spectrum typically behaves as a hard power-law until the maximum energy, which is set by the properties of the accelerator. Most of the energy is here injected around a characteristic energy scale, defined by the maximum proton energy; around this scale, the spectrum looks like a bump.

For this work, we model the diffuse flux as the superposition of two components: a power-law flux, representative of neutrino production in $pp$ sources, and a log-parabola bump-like flux, representative of neutrino production in $p \gamma$ sources (or $pp$ sources with a hard spectrum).  The parametrizations that we adopt for them are flexible enough to capture the variety of behaviors from the interplay of different components.

The diffuse power-law flux component is
\begin{equation}
 \label{equ:pl_flux_def}
 E_\nu^2  \frac{d\Phi_\mathrm{PL}}{dE_\nu dA dt d\Omega}=\Phi_{0,\mathrm{PL}}\left(\frac{E_\nu}{100~{\rm TeV}}\right)^{2-\gamma} e^{-\frac{E_\nu}{E_{\nu,\mathrm{cut}}}} \;,
\end{equation}
where $\Phi_{0, {\rm PL}}$ is a normalization parameter, $\gamma$ is the spectral index, and $E_{\nu, {\rm cut}}$ is the neutrino cut-off energy.  Equation~(\ref{equ:pl_flux_def}) describes the diffuse flux of neutrinos produced in $pp$ interactions of UHECRs that have a relatively soft spectrum $\propto E_p^{-\gamma_p}$ with $\gamma_p \gtrsim 2$, as expected from diffusive shock acceleration.  Below, instead of modeling specific flux predictions, we vary the values of $\Phi_{0, {\rm PL}}$, $\gamma$, and $E_{\nu, {\rm cut}}$ in fits to present-day and projected samples of detected events.

The diffuse bump-like flux component is
\begin{eqnarray}
 \label{equ:bump_flux_def}
 E_\nu^2 
 \frac{d\Phi_{\mathrm{bump}}}
 {dE_\nu dA dt d\Omega}
 &=&
 \left(E_{\nu,\mathrm{bump}}^2 \Phi_{0,\mathrm{bump}}\right)
 \nonumber \\ 
 && \times \exp\left[-\alpha_{\rm bump}\log^2\left(\frac{E_\nu}{E_{\nu,\mathrm{bump}}}\right)\right] \;,
\end{eqnarray}
i.e., a log-parabola, where $E_{\nu, {\rm bump}}^2 \Phi_{0,\mathrm{bump}}$ is a normalization  parameter, $E_{\nu, {\rm bump}}$ is the energy at which the bump is centered, and $\alpha_{\rm bump}$ defines the width of the bump, which is approximately $E_{\nu, {\rm bump}} / \alpha_{\rm bump}^{1/2}$. 
Most of the neutrinos are concentrated around energy $E_{\nu, {\rm bump}}$.  The value of $\alpha_{\rm bump}$ controls whether the spectrum is wide around this energy---if $\alpha_{\rm bump}$ is small---or narrow---if $\alpha_{\rm bump}$ is large.  Equation~(\ref{equ:bump_flux_def}) represents the diffuse flux of neutrinos produced in $p\gamma$ interactions 
(or in $pp$ interactions with a hard CR spectrum).  Below, instead of modeling specific flux predictions, we vary the values of $E_{\nu, {\rm bump}}^2 \Phi_{0,\mathrm{bump}}$, $\alpha_{\rm bump}$, and $E_{\nu, {\rm bump}}$ in fits to present-day and projected samples of detected events.

\section{Hunting for bumps}
\label{sec:hunting_bumps}

We look for bump-like features in the diffuse flux of high-energy neutrinos by using IceCube High-Energy Starting Events (HESE), with high astrophysical purity.  We account for detector effects and the irreducible contamination from  atmospheric neutrinos and muons by using the public IceCube Monte Carlo (MC) HESE sample~\cite{IceCube:2020wum} to compute event rates.

The 7.5-year HESE sample contains 102 events in total.  In our analysis, we use only the 60 events with reconstructed shower deposited energy larger than 60~TeV; there are 41 cascades, 17 tracks, and 2 double cascades.  Above 60~TeV, the contamination from atmospheric neutrinos and muons that pass the HESE self-veto is small, since their fluxes decrease faster with energy than the flux of astrophysical neutrinos.  Because of the event selection, most events are downgoing, namely from the Southern Hemisphere. 

The HESE MC sample contains~821764 simulated HESE events, generated using the same detector simulation used in the analysis of the 7.5-year HESE sample by the IceCube Collaboration. Events in the MC sample were generated assuming a reference diffuse high-energy astrophysical neutrino flux. In our analysis, we compute HESE events corresponding to different choices of the high-energy astrophysical neutrino flux by reweighing the events in the MC sample.  Thus, our predicted event rates inherently include the detailed IceCube HESE response. 

Our statistical procedure is then to use the MC sample to compute the expected event rate for a given astrophysical flux, as defined by the parameters $\boldsymbol \theta \equiv \left( \Phi_{0, {\rm PL}}, \gamma, E_{\nu, {\rm cut}}, E_{\nu, {\rm bump}}^2 \Phi_{0, {\rm bump}},  \alpha_{\rm bump}, \\ E_{\nu, {\rm bump}} \right)$, as defined in Sec.~\ref{sec:pl_bump_components}. The event rate is binned in energy and angle, using the same binning of Ref.~\cite{IceCube:2020wum}. We also account for the atmospheric contamination, leaving the normalization of the different components of the flux free to vary as nuisance parameters, which we collectively defined as $\boldsymbol\eta$; see our main Ref.~\cite{Fiorillo:2022rft} for details. We then define a multi-Poisson likelihood $\mathcal{L}(\boldsymbol\theta,\boldsymbol\eta)$. In terms of this likelihood, we define the evidence
\begin{equation}
 \label{equ:evidence}
 \mathcal{Z}
 =
 \int d\boldsymbol\theta
 \int d\boldsymbol\eta
 ~\mathcal{L}
 \left(
 \boldsymbol \theta, \boldsymbol \eta
 \right)
 \pi\left(\boldsymbol \theta \right)
 \pi\left(\boldsymbol \eta\right) ,
\end{equation}
where $\pi(\boldsymbol\theta)$ and $\pi(\boldsymbol\eta)$ are the prior distributions detailed in our main paper. We evaluate the preference for a two-component, power-law-plus-bump flux model (PLC+B) {\it vs.}~a one-component, power-law flux model (PLC) via the Bayes factor
\begin{equation}
 \label{equ:bayes_factor}
 \mathcal{B}
 =
 \frac{\mathcal{Z}_{\rm PLC+B}}
 {\mathcal{Z}_{\rm PLC}} \;.
\end{equation}
 The higher the value of $\mathcal{B}$, the higher the preference of the data for the two-component flux model.  Broadly stated, narrow bumps are hard to identify---unless they are very tall---because they only affect the event rate within a narrow energy window, while wide bumps are hard to identify because they may resemble a power law.  In-between these extremes, discovery may be more feasible.  We adopt Jeffreys' criteria to classify the preference qualitatively into barely worth mentioning, $10^0 \leq \mathcal{B} < 10^{0.5}$; substantial, $10^{0.5} \leq \mathcal{B} < 10^1$; strong, $10^{1} \leq \mathcal{B} < 10^{1.5}$; very strong, $10^{1.5} \leq \mathcal{B} < 10^2$; and decisive, $\mathcal{B} \geq10^2$. 

Applying this procedure to the present-day, 7.5-year IceCube HESE sample, we find a Bayes factor $\mathcal{B} = 0.7 \pm 0.1$. Following Jeffreys' criteria, this represents mild preference for a one-component power-law flux, to explain the data {\it vs.}~a two-component power-law-plus-bump flux. Although we find evidence against a two-component flux model to explain the present-day HESE data, in what follows we entertain the possibility that instead the present-day best-fit two-component flux is borderline preferred, for two reasons.  First, the present-day preference against the two-component flux is only marginal.  Second, our preference for a two-component flux with a PeV bump is compatible with similar results from previous works~\cite{Palladino:2018evm, Ambrosone:2020evo}, obtained using different methods or event samples and based on specific source models.

\begin{figure*}[t!]
 \centering
 \includegraphics[width=\textwidth]{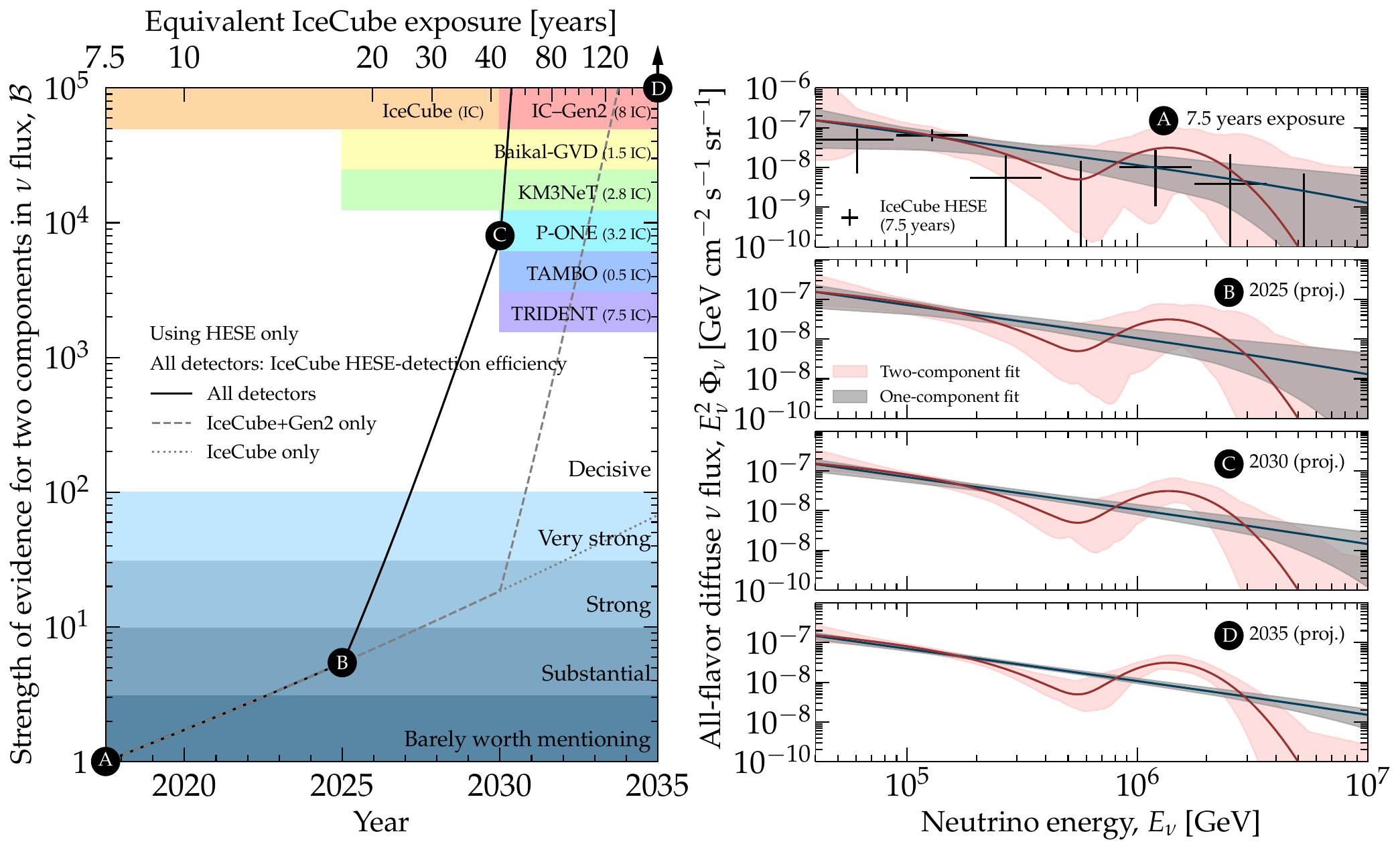}
 \caption{Evidence for the existence of a PeV bump in the diffuse flux of high-energy astrophysical neutrinos.  {\it Left:} Evolution of discovery potential with time, using combined detector exposure. Start times and sizes of upcoming telescopes are estimates for their final configurations (their vertical placements do not convey exposure or evidence).  {\it Right:} Best-fit and 68\% allowed ranges of the one- and two-component flux fits for snapshots A--D. {\it }  {\it A prominent PeV bump may be discovered decisively already by 2027, by combining IceCube, Baikal-GVD, and KM3NeT. Figure taken from Ref.~\cite{Fiorillo:2022rft}.} 
 }
 \label{fig:pev_bump_discovery}
\end{figure*}

Thus, we forecast the discovery prospects of our best-fit two-component flux based on larger HESE event sample made possible by upcoming TeV--PeV neutrino telescopes, currently in operation, construction, and planning stages (see our main paper Ref.~\cite{Fiorillo:2022rft} for a full list of references). Since we do not have simulations of them as detailed as the IceCube one, we model each of them as a re-scaled version of IceCube. Since most of them are in-water or in-ice optical Cherenkov detectors (with the exception of TAMBO, which only detects $\nu_\tau$), this is a reasonable approximation. Of course, it does not capture the differences between detector designs, photomultiplier efficiency, backgrounds, attenuation and scattering length of light in water and ice, systematic errors, and analysis techniques, but it allows us to produce informed estimates of upcoming event rates.

Figure~\ref{fig:pev_bump_discovery} shows the effective volume of each detector, relative to IceCube, and their tentative start dates.  By 2030, we expect nearly an order-of-magnitude increase in the combined detector exposure to high-energy astrophysical neutrinos, thanks to the continuing operation of IceCube and the completion of Baikal-GVD and KM3NeT.  After 2030, we expect a faster growth of the event rate thanks to the construction of new detectors IceCube-Gen2, P-ONE, TAMBO, and TRIDENT.

We also show how the Bayes factor grows with combined detector exposure. Its rate of growth  increases when new detectors are added to the exposure, producing a kink in the slope of the Bayes factor curve.  As expected, because of growing event rates, the longer the exposure, the clearer the separation between the evidence for the one-component and two-component flux fits.  We illustrate the growing separation via four snapshots of the best-fit and 68\% allowed bands of the fluxes, A--D, from present-day to 2035.  

{\it We conclude that the combined exposure of IceCube, Baikal-GVD, and KM3NeT may provide decisive evidence in favor of a two-component flux with a PeV bump already by 2027.}  Alternatively, IceCube plus IceCube-Gen2 may achieve the same by 2031.  {\it In any case, a prominent population of $p\gamma$ sources of PeV neutrinos could be discoverable in the diffuse flux within only a few years.}

\section{Constraining subdominant bumps}

If future observations were to still favor a one-component power-law, we could place upper limits on the height of a coexistent bump component, which must be necessarily subdominant.  We compute the limits as follows.  

\begin{figure}[t!]
 \centering
 \includegraphics[width=0.497\textwidth]{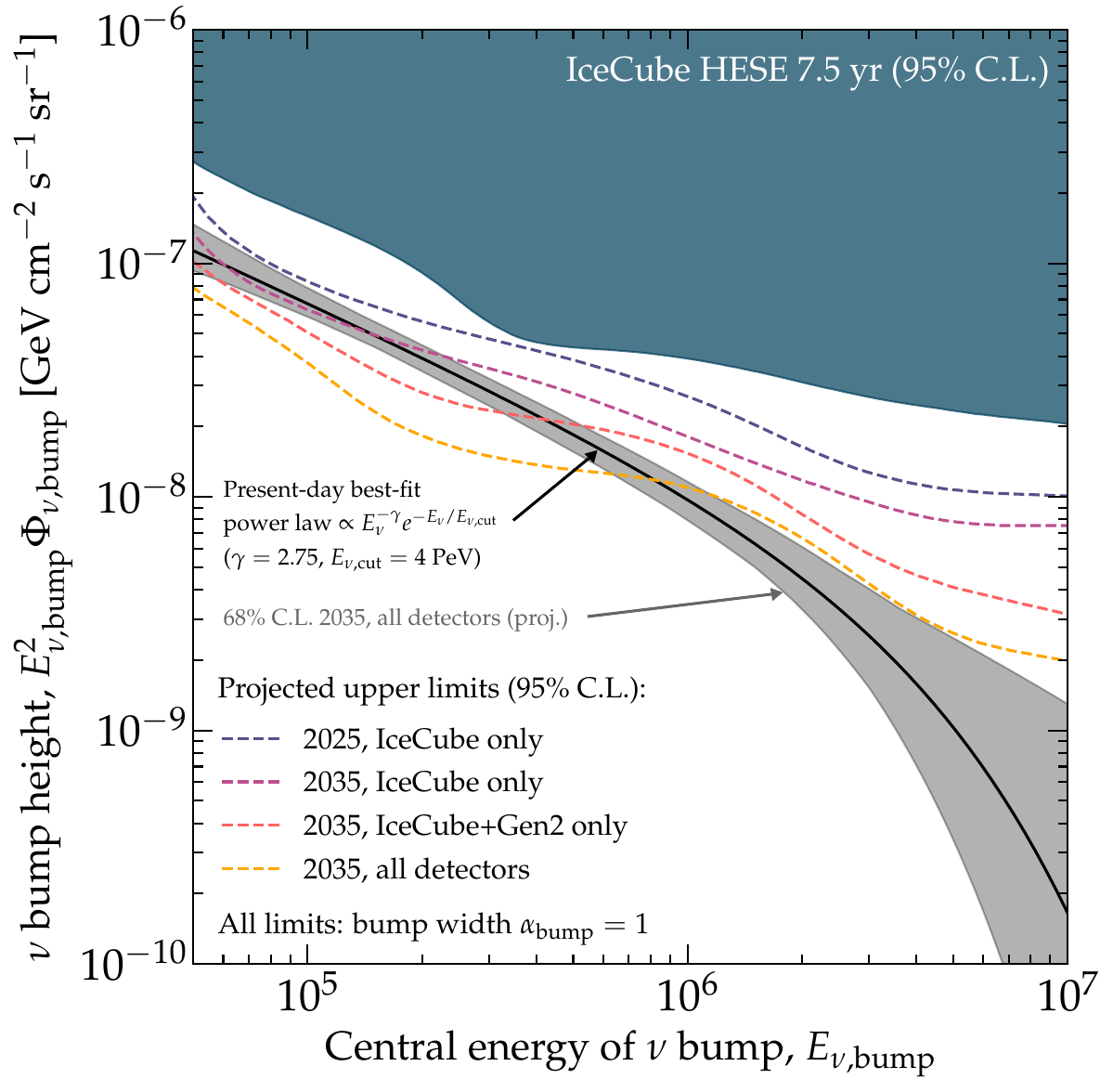}
 \caption{\textbf{\textit{Upper limits on the height of a bump in the diffuse flux of high-energy astrophysical neutrinos.}}  The bump flux component is centered at energy $E_{\nu, {\rm bump}}$, has height $E_{\nu, {\rm bump}}^2 \Phi_{\nu, {\rm bump}}$, and width $\alpha_{\rm bump} = 1$, and is overlaid on a power-law flux $\propto E_\nu^{-\gamma} e^{E_\nu/E_{\nu, {\rm cut}}}$, with parameter values given by the best fit to the 7.5-year IceCube HESE sample~\cite{IceCube:2020wum}, shown for comparison.  \textit{Today, IceCube limits the height of a bump centered at a few hundred TeV to be, at most, comparable to the size of the dominant power-law component.  In the future, the upper limit may be tightened to tens of percent of the power-law component. Figure taken from Ref.~\cite{Fiorillo:2022rft}.}  
 }
 \label{fig:constraints_alpha_bump_1}
\end{figure}

For given values of the position of the bump, $E_{\nu,\mathrm{bump}}$ and of its width, which we keep fixed at the representative value of $\alpha_{\rm bump} = 1$ in the main text, we compute the posterior under the two-component flux model and marginalize it over all the free model parameters, except for the bump height, $E^2_{\nu,\mathrm{bump}}\Phi_{0,\mathrm{bump}}$.  We integrate the resulting one-dimensional marginalized posterior to find the 95\% credible interval on the bump height, for each value of the bump position.  Differently from our previous results, in drawing constraints on the bump height we adopt a flat linear prior on it, rather than a logarithmic one.  (Otherwise, because the posterior is flat for arbitrarily low values of the bump height, limits drawn using a logarithmic prior would differ depending on our arbitrary choice of the lower end of the logarithmic prior.)

Figure~\ref{fig:constraints_alpha_bump_1} shows the results.  Present-day limits, based on the 7.5-year IceCube HESE sample, disfavor especially the presence of relatively wide bumps, with $\alpha_{\rm bump} = 1$, centered around 200~TeV, where event statistics are higher. We choose $\alpha_{\rm bump} = 1$ as a benchmark value that lies between the two extremes of a very wide bump, with $\alpha_{\rm bump} \ll 2$, and a very narrow bump, with $\alpha_{\rm bump} \gg 2$.  For all values of $E_{\nu, {\rm bump}}$, the limit lies above the present-day best-fit power-law flux, meaning that a sizable contribution to the diffuse flux from a population of photohadronic sources cannot presently be excluded.  The limits are weaker for bumps centered at lower energies, where the atmospheric background is higher, and at higher energies, where statistics are poorer.  The weakening above $500$~TeV reflects the fact that a two-component flux with a bump between hundreds of TeV and a few PeV is only marginally disfavored in present-day data.

\section{Discussion}

We have performed a systematic search in the energy spectrum of present-day IceCube data and make forecasts based on expected future data.  We look for features in the diffuse spectrum from two broad source classes: sources that make neutrinos via proton-proton ($pp$) interactions or photohadronic, i.e., proton-photon ($p\gamma$) interactions. We adopt flexible spectral shapes for the power-law and bump-like fluxes, that allow us to probe many different shapes and relative sizes of them. To deliver on the full potential of our methods, we extend our analysis to the expected combined exposure of multiple upcoming neutrino detectors.

Present-day data do not show a clear evidence in favor of a multi-component flux -- in fact, a single power-law flux, being most economical in terms of parameters space, is the preferred explanation. However, this is unsurprising, given the small statistics that we have available. Therefore, we crucially extend our analysis to the combined exposure of the upcoming detectors. Our results reveal that, if indeed the diffuse neutrino flux is composed of a power law and a bump component, conclusive evidence for the features deriving from the transition between the two might show up within a few years of exposure of the combined detectors.

On the other hand, if no bump-like component is identified, the increased exposure available in the next years will allow to set powerful constraints on how large its contribution can be to the diffuse flux. We have outlined a strategy to do so, and shown that by 2035 the combined exposure of all detectors will allow to constrain a bump contribution to a fraction of the diffuse flux in the high-statistics region. We emphasize that these bounds are entirely independent from the similar ones obtained using point-source searches; they are based on purely spectral information, and can therefore complement the point source ones.

\end{document}